\documentclass[journal=jpcafh,manuscript=article]{achemso}

\usepackage{chemformula}
\usepackage[T1]{fontenc}

\usepackage[sort&compress,numbers]{natbib}
\setkeys{acs}{doi = true}
\usepackage{doi}

\usepackage{hyperref}
\hypersetup{colorlinks=true,linkcolor=blue,anchorcolor=blue,
            citecolor=blue,filecolor=blue,urlcolor=blue}

\author{Jean-Nicolas Vigneau}
\affiliation{Université Paris-Saclay, CNRS, Institut des Sciences Moléculaires d’Orsay\\ 91405 Orsay cedex, France}
\alsoaffiliation{Département de chimie, COPL, Université Laval, 1045 av. de la Médecine\\ Québec, QC G1V 0A6, Canada}

\author{Thanh-Tung Nguyen Dang}
\affiliation{Département de chimie, COPL, Université Laval, 1045 av. de la Médecine\\ Québec, QC G1V 0A6, Canada}

\author{Eric Charron}
\email{eric.charron@universite-paris-saclay.fr}
\affiliation{Université Paris-Saclay, CNRS, Institut des Sciences Moléculaires d’Orsay\\ 91405 Orsay cedex, France}

\title[Dynamics of \ch{H2}]{Electro-nuclear dynamics of single and double ionization of \ch{H2} in ultrafast intense laser pulses}

\begin{document}

\begin{abstract}
We present an efficient method for modeling the single and double ionization dynamics of the \ch{H2} molecule in ultrashort intense laser fields. This method is based on a semi-analytical approach to calculate the time-dependent single and double molecular ionization rates and on a numerical approach to describe the vibrational motion that takes place in the intermediate molecular ion \ch{H2+}. This model allows for the prediction of the single and double ionization probabilities of the \ch{H2} molecule to be made over a wide range of frequencies and laser intensities with limited computational time, while providing a realistic estimate of the energy of the products of the dissociative ionization and of the Coulomb explosion of the \ch{H2} molecule. The effect of vibrational dynamics on ionization yields and proton kinetic energy release spectra is demonstrated and, in the case of the latter, discussed in terms of basic strong-field molecular fragmentation mechanisms.
\end{abstract}

\section{Introduction}
The interaction of gas-phase molecules with intense ultrashort laser fields can often lead to the coupling of electronic and nuclear dynamics, especially in cases where the field is intense enough to induce multiple ionization phenomena. The prototype molecule for exploring these phenomena over the last 30 years has been the \ch{H2} {molecule\cite{Giusti_1995, Sheehy_1996, Posthumus_2004}}. The hydrogen molecule is unique in that its nuclear dynamics unfolds rapidly, on a timescale of just a few femtoseconds, inducing an entanglement of the electronic and nuclear dynamics that is not always easy to decipher. It is also one of the simplest molecules, with a limited number of electronic and nuclear degrees of freedom, for which a variety of realistic and accurate theoretical models have been {developed\cite{Giusti_1995, Palacios_2015}}. It is therefore clearly a prototype molecule that has allowed the discovery of various specific and sometimes unexpected molecular mechanisms at work when a molecule is subjected to intense laser fields: Above Threshold {Dissociation\cite{Giusti-Suzor1990, Bucksbaum1990, Zavriyev1990, Yang1991}}, Bond {Softening\cite{Bucksbaum1990, Zavriyev1990, Yang1991, Jolicard1992, Atabek_1994}}, Vibrational Trapping also known as Bond {Hardening\cite{Giusti_1992, Yao_1993, Zavriyev_1993, Aubanel_1993, Aubanel_1993b}}, Coulomb {Explosion\cite{Codling_1988, Codling_1993}}, and Charge Resonance Enhanced {Ionization\cite{Zuo_1995, Constant_1996}}. All of these fundamental molecular fragmentation mechanisms were discovered and studied in the last decade of the 20th century. At that time, laser pulses with a duration comparable to the vibrational period of \ch{H2} were not yet available to researchers. It is only with the recent development of new light sources that it became possible to use pulses comprising just a few optical cycles, {i.e.\!} of slightly less than ten femtoseconds at the carrier wavelength of {800\;nm\cite{Corkum_2007}}. Thanks to these recent developments, new studies have been carried out with the aim of elucidating in real time the different dynamics at work in the hydrogen molecule fragmentation processes in intense laser {fields\cite{Ergler_2005, Legare_2005, Ergler_2006, Ergler_2006b, Niikura_2006, Rudenko_2007, Bocharova_2008, Calvert_2010, Palacios2014, Lu_2018}}.

From a theoretical point of view, although \ch{H2} is one of the simplest molecules we can consider, the accurate modeling of its interaction with an intense laser field, leading to its dissociative ionizations up to its complete Coulomb explosion, remains a real challenge due to the large number of degrees of freedom to be dealt with, while taking into account the quantum dynamics that unfold on very different time scales for the motion of the electrons (in the attosecond range) and of the nuclei (in the femtosecond range). As a result, many theoretical studies have been performed using simplified models that can only be used in certain limited ranges of laser parameters. Examples include ionization dynamics studies performed with the position of the nuclei {fixed\cite{Giusti_1995, Zuo_1995}}, photodissociation dynamics studies that do not consider ionization {processes\cite{Giusti-Suzor1990, Giusti_1992, Giusti_1995}}, or studies performed in reduced {dimensionality\cite{Saugout2007, Saugout2008, Lu_2018, Vigneau2022}}. These types of approaches, which have proven to be very interesting in order to better analyze, understand and sometimes control the processes that take place during or after laser excitation, often impose severe constraints on the conditions under which these models can be applied, for example on the intensity range in which they can realistically be used. While it is true that much more sophisticated models have been developed recently, such as the time-dependent Feshbach close-coupling {approach\cite{Palacios_2015}}, there are applications where it would be extremely useful to have approximate models that could estimate, with reduced computational times, the single and double ionization probabilities of \ch{H2}\,, and at the same time predict the energy of the protons emitted during the fragmentation of the molecule.

This is especially the case in particle-in-cell (PIC) codes, which are widely used in laser-plasma interaction physics. In these systems, the \ch{H2} molecule has recently gained particular interest, as it has been used to demonstrate laser wakefield acceleration of quasi-monoenergetic electron bunches up to very high energies with high repetition {rates\cite{Salehi_2021}}. For the calculation of single and double ionization processes, the PIC codes currently in use are based on simple Ammosov-Delone-Krainov (ADK) atomic models\cite{ammosov_1986} with ionization potentials adapted to the \ch{H2} {molecule\cite{Salehi_2021, Maldonado_2019}}. These approaches are very approximate and cannot predict the energy of the emitted protons, which is a variable that can influence plasma dynamics. 

\begin{figure}[!t]
\centering
\includegraphics[height=12cm]{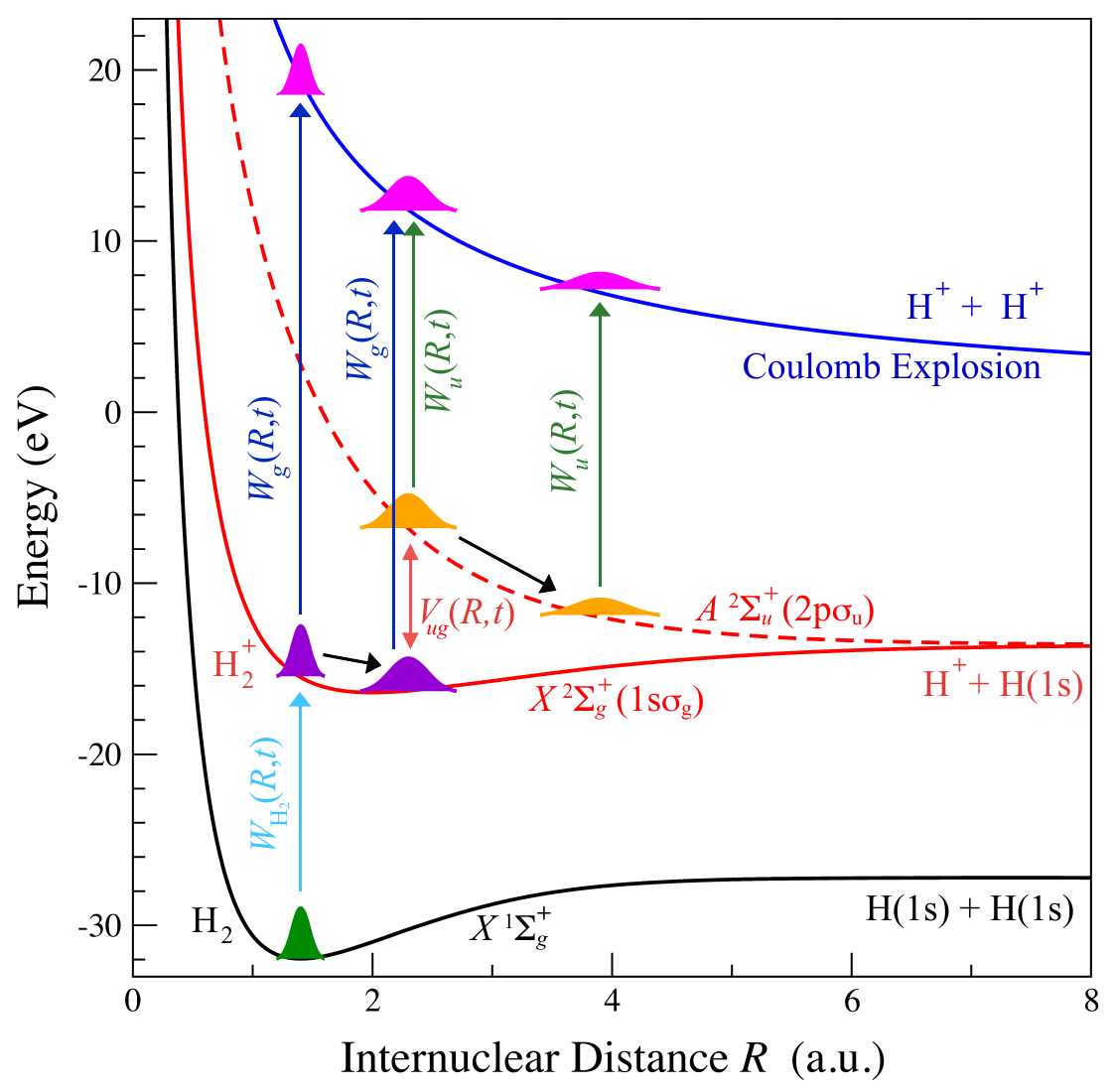}
\caption{Representation of the first and second ionization paths of the \ch{H2} molecule, initially in its electronic and vibrational ground states. The lowest energy potential curve, shown in black, corresponds to the $X^1\Sigma_g^+$ electronic state of \ch{H2}. The two red potential curves, solid and dashed, correspond to the two lowest energy electronic states, $X^2\Sigma_g^+$ (1s$\sigma_g$) and $A^2\Sigma_u^+$ (2p$\sigma_u$), of the \ch{H2+} molecular ion. Finally, the highest energy potential curve, shown in blue, corresponds to the Coulomb repulsion of the doubly ionized \ch{H+}~\!+~\ch{H+} system. The initial $v=0$ wave function of \ch{H2} is shown in green. Single ionization generates a vibrational wave packet, shown in purple, on the ground electronic curve of the \ch{H2+} molecular ion. This wave packet evolves while coupled to the first dissociative excited state of \ch{H2+}, giving rise to a second wave packet, shown in orange. The nuclear dynamics of these two wave packets in the 1s$\sigma_g$ and 2p$\sigma_u$ electronic states is accompanied by a possible second ionization, which ultimately leads to the Coulomb explosion process, schematically represented here by the three pink wave packets. $W_{\ch{H2}}(R,t)$ is the instantaneous single ionization rate of \ch{H2} at internuclear distance $R$ and time $t$. Similarly, $W_g(R,t)$ and $W_u(R,t)$ are the instantaneous ionizations rates of \ch{H2+} in the lowest 1s$\sigma_g$ and highest 2p$\sigma_u$ electronic states. Finally, $V_{ug}(R,t)$ is the radiative coupling between the 1s$\sigma_g$ and 2p$\sigma_u$ states at distance $R$ and time $t$.}
\label{Fig:Pot_Curves}
\end{figure}

In the model proposed here, we have chosen to treat single and double ionization phenomena using the  molecular version of the quasi-analytical {Perelemov-Popov-Terent'ev\cite{PPT_1, PPT_2, PPT_3, Tong2002, Benis2004}} (PPT) expressions for the ionization rates and to combine it with nuclear wave packet {propagation\cite{Charron1995}}. This combination of an analytical result derived from a semi-classical quasi-static theory of ionization with fully quantum-mechanical time-dependent wave packet calculations is what makes the present approach original. It is  aimed at taking into account the nuclear dynamics that is induced by the different excitation and ionization processes. This allows us to consider, in a fairly simple yet realistic way, the ionization of \ch{H2} initially in its ground electronic state, while taking into account the possible stretching of the molecule. Once \ch{H2+} is formed in its ground electronic state, its nuclear wave packet is propagated on this state coupled to the first electronic state of the ion. The two wave packets thus formed in \ch{H2+} can further ionize, giving rise to the Coulomb explosion process. The excitation pathways followed by the molecule are shown schematically in Fig.\;\ref{Fig:Pot_Curves}. This model allows one to study in detail the influence of vibrational dynamics on single and double ionization processes in the \ch{H2} molecule, while covering a wide range of laser intensities, from relatively low intensities where above threshold dissociation and bond softening processes dominate, to higher intensities where Coulomb explosion becomes the dominant process. It also allows to cover different wavelength ranges and to predict the energy of the emitted protons with a reasonable computational effort.

In the following sections, we first introduce the theoretical model of strong field photoionization and photodissociation dynamics of molecular hydrogen used in this study. Then, we present the influence of vibrational dynamics on ultrashort single-cycle and multi-cycle pulses at different wavelengths and intensities, and finally, we discuss proton kinetic energy release spectra calculated at different intensities. To conclude this introduction, we would like to mention that this work would not have been possible without the invaluable help of our dear late colleague Osman Atabek (April 28, 1946 - June 27, 2022), whose memory we would like to honor with this publication.

\section{Methods}

In this time-dependent approach, to account for the nuclear dynamics occurring during a single or double ionization event induced by a strong linearly polarized laser pulse, the molecular system is described by the nuclear wave packets $\Psi_{\ch{H2}}(R,t)$, $\Psi_g(R,t)$, $\Psi_u(R,t)$ and $\Psi_{\mathrm{C}}(R,t)$, which represent the neutral molecule \ch{H2} in its ground electronic state, the molecular ion \ch{H2+} in its two lowest electronic states 1s$\sigma_g$ and 2p$\sigma_u$, and the doubly charged ion \ch{H2^{2+}} in the Coulomb explosion channel, respectively. The \ch{H2} molecule is assumed to be initially in its ground vibrational state, with the nuclear wave function
\begin{equation}
\Psi_{\ch{H2}}(R,0) = \chi_0(R)\,, 
\end{equation}
and with $\Psi_g(R,0) = \Psi_u(R,0) = \Psi_{\mathrm{C}}(R,0) = 0$. The ground vibrational wavefunction $\chi_0(R)$ is
obtained by applying Numerov's approach \cite{Numerov1924, Numerov1927}  to the potential energy function of the ground electronic state of {\ch{H2} \cite{Kolos2004}}. The evolution over a short time interval (of extension $\delta t$) of the nuclear wave packet $\Psi_{\ch{H2}}(R,t)$ representing the \ch{H2} molecule is then performed numerically in two successive steps. In the first step, the population loss due to the ionization of the neutral molecule is taken into account by introducing the instantaneous ionization rate $W_{\ch{H2}}(R,t)$, according to
\begin{equation}
\overline{\Psi}_{\ch{H2}}(R, t) = \Big[\, 1 - W_{\ch{H2}}(R,t)\, \delta t \,\Big]^{\frac{1}{2}}\; \Psi_{\ch{H2}}(R,t)\,.
\label{Eq:H2_ionization}
\end{equation}
To evaluate the ionization rate of \ch{H2} in an intense laser field, we use the molecular ionization {model \cite{Tong2002, Benis2004}} derived from the PPT {approach \cite{PPT_1, PPT_2, PPT_3}}, as described in detail {in \citet{Vigneau2023}} for instance. Finally, in a second step, the resulting wave packet $\overline{\Psi}_{\ch{H2}}(R, t)$ is propagated numerically according to
\begin{equation}
\Psi_{\ch{H2}}(R, t + \delta t) = \hat{U}_{\ch{H2}}(t + \delta t \leftarrow t)\, \overline{\Psi}_{\ch{H2}}(R, t)\,,
\label{Eq:H2_propagation}
\end{equation}
with the evolution operator $\hat{U}_{\ch{H2}}(t + \delta t \leftarrow t)$ evaluated using the split operator {method \cite{Feit1982}}, which optimally splits into three factors associated with the nuclear kinetic energy and potential energy {operators \cite{Split-Bandrauk1993}}. The passage from one to another of these non-commuting factors makes use of Fast Fourier {Transforms\cite{Feit1982}}.

From Eq.\,(\ref{Eq:H2_ionization}) we can deduce that the variation of the population in \ch{H2} between time $t$ and time $t + \delta t$ is equal to
\begin{equation}
\Delta P_{\ch{H2}}(t + \delta t \leftarrow t) = - \delta t \int W_{\ch{H2}}(R,t)\, \left|\Psi_{\ch{H2}}(R,t)\right|^2\,dR\,,
\label{Eq:H2_loss}
\end{equation}
where the minus sign indicates a population loss. It is well known that molecular ionization rates can vary significantly with $R$, since the ionization potential itself is usually a function of the internuclear {distance \cite{Vigneau2023}}. We can see in Eq.\;(\ref{Eq:H2_loss}) that this important dependence of the ionization rate on $R$ is well accounted for in this model.

This population loss in \ch{H2} essentially populates \ch{H2+} in its $1s\sigma_g$ electronic ground {state \cite{Vigneau2023}}. Similar to what is done for \ch{H2}, the evolution of the $\Psi_g(R,t)$ and $\Psi_u(R,t)$ wave packets representing \ch{H2+} is carried out in successive steps. In the first step, the population gain due to the ionization of the neutral molecule in the \ch{H2+} ground state is calculated using
\begin{equation}
\overline{\Psi}_g(R,t) = \Psi_g(R,t) -i\, \alpha(t)\, \Big[\, W_{\ch{H2}}(R,t)\, \delta t \,\Big]^{\frac{1}{2}}\; \Psi_{\ch{H2}}(R,t)\,,
\label{Eq:H2+_gain}
\end{equation}
where $\alpha(t)$ is a normalization factor calculated numerically to impose a population gain in \ch{H2+} that exactly offsets, during each time step $\delta t$, the loss in \ch{H2} given in Eq.\;(\ref{Eq:H2_loss}). The nuclear wave packet associated with the first excited electronic state of \ch{H2+} remains unchanged at this point, with
\begin{equation}
\overline{\Psi}_u(R,t)=\Psi_u(R,t)\,,
\end{equation}
since the ionization of \ch{H2} in an intense laser field produces \ch{H2+} in its ground electronic state {only \cite{Vigneau2023}}. In a second step, the two coupled wave packets $\overline{\Psi}_g(R,t)$ and $\overline{\Psi}_u(R,t)$ are propagated numerically according to
\begin{equation}
\left[
\begin{array}{c}
\overline{\overline{\Psi}}_g(R,t + \delta t)\\
\overline{\overline{\Psi}}_u(R,t + \delta t)
\end{array}
\right]
= \hat{U}_{\ch{H2+}}(t + \delta t \leftarrow t)
\left[
\begin{array}{c}
\overline{\Psi}_g(R,t)\\
\overline{\Psi}_u(R,t)
\end{array}
\right],
\label{Eq:H2+_propagation}
\end{equation}
where the evolution operator is evaluated as
\begin{equation}
\hat{U}_{\ch{H2+}}(t + \delta t \leftarrow t) = 
\exp{\left[-i\;\frac{\hbar\delta t}{4m}\frac{\partial^2~}{\partial R^2}\,\mathbf{I}\right]}\;
\exp{\left[-i\;\frac{\delta t}{\hbar}\,\mathbf{V}(R,t)\right]}\;
\exp{\left[-i\;\frac{\hbar\delta t}{4m}\frac{\partial^2~}{\partial R^2}\,\mathbf{I}\right]}\;
\label{Eq:evolutionO}
\end{equation}
to decrease the error to the order $(\delta t)^3$. In this equation, $m$ is the nuclear reduced mass associated with \ch{H2+}, $\mathbf{I}$ is the $2 \times 2$ identity matrix, and $\mathbf{V}(R,t)$ is the $2 \times 2$ potential matrix composed on its diagonal by the electronic potentials $V_g(R)$ and $V_u(R)$ associated with the 1s$\sigma_g$ and 2p$\sigma_u$ states, and by the $V_{ug}(R,t)=-\mu_{ug}(R)E(t)$ non-diagonal radiative coupling term
\begin{equation}
\mathbf{V}(R,t) =
\begin{bmatrix}
V_g(R)    & V_{ug}(R,t) \\
V_{ug}(R,t) & V_u(R)
\end{bmatrix}\;.
\end{equation}
$\mu_{ug}(R)$ is the transition moment between the 1s$\sigma_g$ and 2p$\sigma_u$ states, taken from \citet{Charron1995}. The kinetic propagation is performed in momentum space and the potential propagation is performed in coordinate space using the diagonal representation of the potential matrix $\mathbf{V}(R,t)$. The evolution operator (\ref{Eq:evolutionO}) thus allows to describe between the times $t$ and $t + \delta t$ the nuclear dynamics induced in \ch{H2+} by the kinetic and potential energy of the nuclei, while introducing on an equal footing the laser coupling between the two electronic states $1s\sigma_g$ and $2p\sigma_u$ of \ch{H2+}. It is therefore during this coupled propagation step that the first (2p$\sigma_u$) electronic state of \ch{H2+} is populated. Finally, in a third and last step, the population losses due to the ionization of \ch{H2+} are taken into account using the molecular-PPT \cite{Vigneau2023} ionization rates $W_{g}(R,t)$ and $W_{u}(R,t)$ associated with the $1s\sigma_g$ and $2p\sigma_u$ states
\begin{subequations}
\begin{eqnarray}
\Psi_g(R,t + \delta t) & = & \Big[\, 1 - W_{g}(R,t + \delta t)\, \delta t \,\Big]^{\frac{1}{2}}\; \overline{\overline{\Psi}}_g(R,t + \delta t)\\
\Psi_u(R,t + \delta t) & = & \Big[\, 1 - W_{u}(R,t + \delta t)\, \delta t \,\Big]^{\frac{1}{2}}\; \overline{\overline{\Psi}}_u(R,t + \delta t)\,.
\end{eqnarray}
\label{Eq:H2+_ionization}
\end{subequations}
These ionization events, which strip the molecule of its last electron, give rise to a Coulomb explosion phenomenon. From Eqs.\;(\ref{Eq:H2+_ionization}) we can derive the vibrational distribution formed in the Coulomb explosion channel $P_{\mathrm{vib}}(R,t + \delta t \leftarrow t)$ between times $t$ and $t + \delta t$ as
\begin{equation}
P_{\mathrm{vib}}(R,t + \delta t \leftarrow t) = \left[\,W_{g}(R,t)\,\left|\overline{\overline{\Psi}}_g(R,t)\right|^2 + W_{u}(R,t)\,\left|\overline{\overline{\Psi}}_u(R,t)\right|^2\,\right] \delta t\,.
\end{equation}
Note that this is an incoherent sum of probabilities (corresponding to two orthogonal components of $\Psi_c(R,t)$ not explicitly given here), since the electrons emitted from the 1s$\sigma_g$ and 2p$\sigma_u$ orbitals are characterized by different symmetries. Due to the simple $1/R$ law followed by the Coulomb repulsion energy, a simple mapping relates this vibrational probability distribution to the kinetic energy distribution of the emitted protons. This relation is a consequence of the conservation principle \cite{Saugout2007, Saugout2008}
\begin{equation}
P_{\mathrm{c}}(E,t + \delta t \leftarrow t)\,dE = P_{\mathrm{vib}}(R,t + \delta t \leftarrow t)\,dR
\end{equation}
which leads to
\begin{equation}
P_{\mathrm{c}}(E,t + \delta t \leftarrow t) = 2R^2 \left[\,W_{g}(R,t)\,\left|\overline{\overline{\Psi}}_g(R,t)\right|^2 + W_{u}(R,t)\,\left|\overline{\overline{\Psi}}_u(R,t)\right|^2\,\right] \delta t
\end{equation}
in atomic {units\cite{Saugout2008}}. This energy distribution is accumulated over the entire time propagation. The result is the kinetic energy release spectrum $P_{\mathrm{c}}(E)$, which is measured in experiments where Coulomb explosion is induced by a strong laser field.

To complete this kinetic energy release spectrum of protons produced by Coulomb explosion, we need of course to add the spectrum resulting from the photo-dissociation of the molecular ion \ch{H2+}. This additional spectrum is calculated by projecting the nuclear wave packets $\Psi_g(R,t_{\!f})$ and $\Psi_u(R,t_{\!f})$ obtained at the end of the pulse $(t=t_{\!f})$ onto the energy-normalized solutions of the field-free dissociated molecular states, as explained in detail e.g. {in \citet{Charron1995}}. This gives access to the calculation of a complete proton kinetic energy release spectrum $P(E)$, including both the low-energy part associated with the photo-dissociation of \ch{H2+} leading to \ch{H+} + H(1s) fragments, and the high-energy part associated with the Coulomb explosion channel leading to \ch{H+} + \ch{H+} fragments.

Finally, the \ch{H2} and \ch{H2+} populations at each time $t$ are obtained by calculating the norms of the wave packets $\Psi_{\ch{H2}}(R,t)$, $\Psi_g(R,t)$ and $\Psi_u(R,t)$, while the population associated with the Coulomb explosion channel can be calculated by accumulating the population losses of the 1s$\sigma_g$ and 2p$\sigma_u$ states over the course of the pulse, or by integrating the kinetic energy spectrum $P_{\mathrm{c}}(E)$ associated with Coulomb explosion.

\section{Results and Discussion}

In this section, starting with the case of ultrashort single-cycle pulses, we first evaluate the importance of considering vibrational dynamics in the calculation of proton kinetic energy release (KER) spectra. We then analyze the effect of the vibrational dynamics on the single and double ionization probabilities for pulses of a few optical cycles at 800 nm and at different peak intensities. Finally, we discuss and interpret proton KER spectra calculated at 266\;nm for different laser intensities. 

From a practical point of view, simulations are performed with a time step $\delta t$ equal to one thousandth of the optical cycle duration. Thus, at 266\;nm this time step is $\delta t \simeq 0.9$\;as, while at 800\;nm it is about 2.7\;as. All results presented here assume a linear polarization of the field and a temporal envelope of the pulse defined by a $\sin^2$ function \footnote{This choice is made over that of an experimentally more relevant Gaussian envelope because the two are extremely close in value around the peak of the intense laser pulse, where most excitation processes, in particular tunnel ionization, occur. In addition, the $\sin^2$ envelope has the advantage of having a well-defined beginning and end at finite times, where the system is in a truly field-free condition.}.
The time-dependent electric field is therefore written as
\begin{equation}
E(t) = E_0\, \sin^2\!\left(\frac{\pi\,t}{t_{\!f}}\right) \cos(\omega t)
\end{equation}
where $E_0$ is the electric field amplitude and $\omega$ the laser carrier frequency.

In theoretical studies dealing with the ionization of \ch{H2} in an intense laser field, it is common to invoke the the small mass of the electrons relative to the mass of the nuclei to get rid of the motion of the nuclei by modeling the electronic dynamics at a fixed inter-nuclear {distance \cite{Vigneau2022}}. This approximation is chosen mainly for numerical reasons, since a complete quantum calculation including all degrees of freedom of the system, both electronic and nuclear, is computationally extremely expensive. We nevertheless expect that for ultrashort single-cycle pulses, the impact of vibrational dynamics remains limited, even for a molecule as light as \ch{H2}. Indeed, in its ground electronic state, \ch{H2} is characterized by a vibrational period of about 7.5\,fs, which is clearly longer than the duration of an optical cycle in the range between the near UV and the near IR.

\begin{figure}[!t]
\centering
\includegraphics[height=11cm]{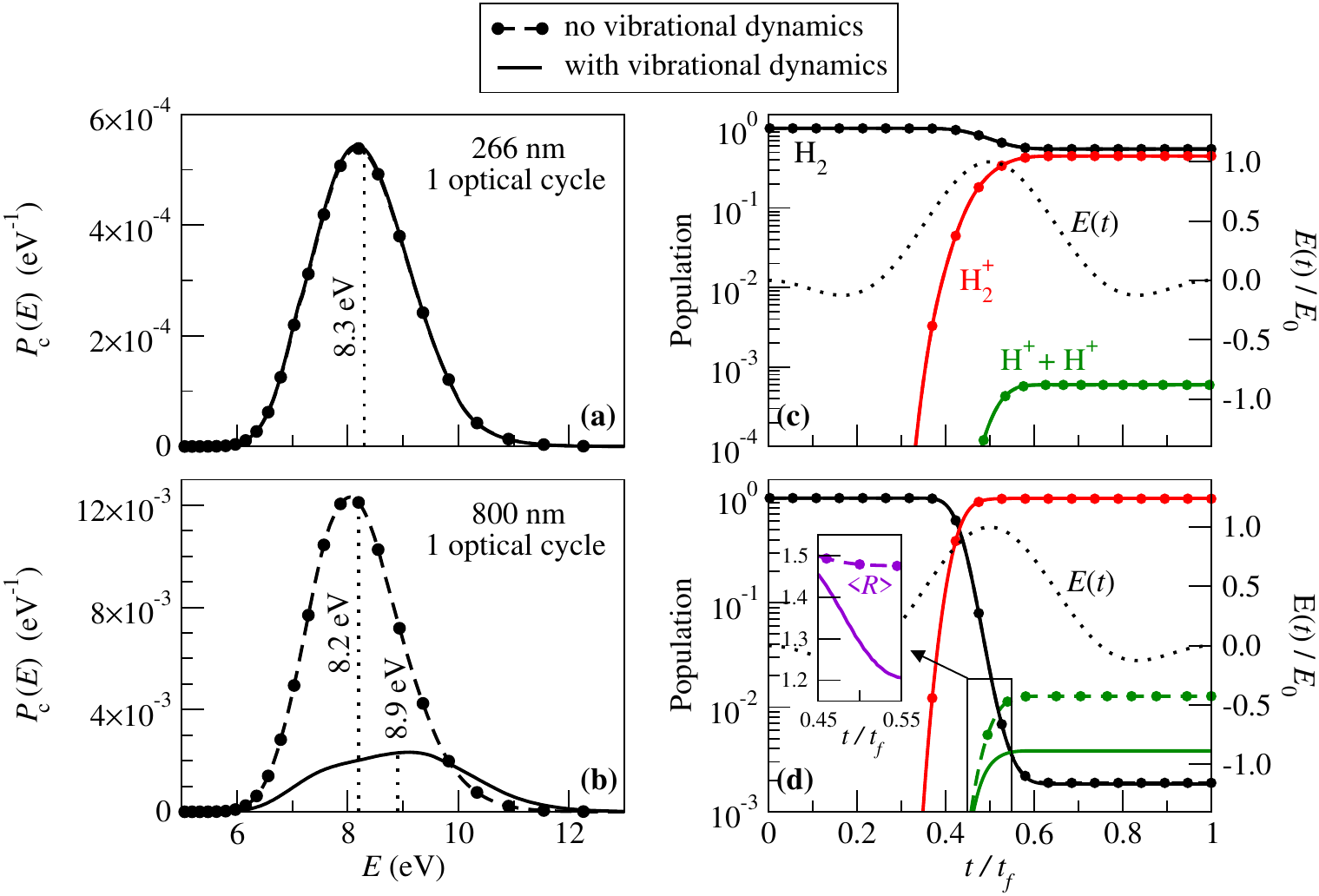}
\caption{Coulomb explosion spectra and ionization dynamics for one-optical-cycle pulses at 266\,nm and 800\,nm with a peak intensity of $I=10^{15}$\,W/cm$^2$. The first row (panels \textbf{a} and \textbf{c}) corresponds to 266\,nm, while the second is for 800\,nm. The dashed curves with circles are obtained by freezing the nuclear dynamics, while the solid curves take the nuclear dynamics into account. The left side of the figure (panels \textbf{a} and \textbf{b}) shows the Coulomb explosion spectra $P_{\mathrm{c}}(E)$. The right side of the figure (panels \textbf{c} and \textbf{d}) shows the time evolution of the \ch{H2} populations in black, \ch{H2+} in red and \ch{H+}~\!+~\ch{H+} in green on a logarithmic scale (left axis). The inset in panel \textbf{d} shows the time evolution of the average internuclear distance $\langle R \rangle$ (in atomic units) of the \ch{H2+} 1s$\sigma_g$ wave packet in the case where the nuclear dynamics is frozen (purple dashed line with circles) and in the case where it is not (purple solid line). The dotted black lines in panels \textbf{c} and \textbf{d} show the time dependence of the normalized electric field $E(t)/E_0$ on a linear scale (right axis).}
\label{Fig:One_Cycle}
\end{figure}

To highlight the influence of vibrational dynamics in the case of ultrashort pulses, we show in panels \textbf{a} and \textbf{b} (left column) of Fig.\;\ref{Fig:One_Cycle} the Coulomb explosion KER spectra $P_{\mathrm{c}}(E)$ calculated for a single optical cycle pulse at $10^{15}$\;W/cm$^2$ in the near UV, specifically at 266\,nm (panel \textbf{a}) and in the near IR, at 800\,nm (panel \textbf{b}). The results obtained by freezing the vibrational dynamics are shown in dashed lines with circles, while the solid curves are associated with the complete calculation taking into account the motion of the nuclei.

At 266\;nm there is almost no difference between the two curves, which are superimposed on top of each other, indicating that taking the vibrational dynamics into account is not necessary in the case of a pulse with a total duration of only 0.9\;fs. Note that the calculated Coulomb explosion peak, slightly asymmetric, is centered on an average kinetic energy of about 8.3\;eV per proton. This average energy cannot be derived from a simple projection of the initial wave function of \ch{H2} onto the Coulomb energy curve $E=1/R$, since this simplistic approach would give, for an average internuclear distance in \ch{H2} of $R_{eq}\simeq 1.4$\;a.u., an average energy of 9.7\;eV. The observed shift to a lower energy (8.3\;eV) is due to the fact that the  rates of first ionization, $W_{\ch{H2}}(R,t)$, and of second ionization, $W_g(R,t)$, increase significantly with the internuclear distance\cite{Vigneau2023} due to a gradual decrease of the respective molecular ionization potentials with the internuclear distance. This property of the ionization rates therefore tends to favor the ionization of the part of the initial wavefunction located at $R > R_{eq}$ to the detriment of the part at $R < R_{eq}$, thus inducing a Coulomb explosion peak at a lower energy than what would have been estimated in a purely vertical excitation scheme.

Continuing with the case of a $266$\;nm pulse, the panel \textbf{c} in Fig.\;\ref{Fig:One_Cycle} shows the population dynamics during that pulse in a logarithmic scale. The populations in \ch{H2}, \ch{H2+} and in the Coulomb explosion channel are shown in black, red and green respectively. Here also, the results of the calculations with and without vibrational dynamics are superimposed. The time evolution of the electric field is also shown schematically on this graph as a black dotted line. We can see a single burst of single and double ionization, around $t = t_f/2$, i.e. at the time when the electric field is at its maximum, delivering a peak intensity of $10^{15}$\;W/cm$^2$.

If we now look in panel \textbf{b} of Fig.\;\ref{Fig:One_Cycle} at the influence of the vibrational dynamics on the Coulomb explosion spectrum at 800\,nm in the case of a single optical cycle, the contrast with what we have just discussed at 266\;nm is clearly visible. Indeed, the KER spectra calculated with and without vibrational dynamics are quite different. This is due to the fact that at 800\;nm the total duration of the single-cycle pulse is 2.7\;fs. This duration is not negligible compared to the characteristic time associated with the motion of the nuclei (7.5\,fs). An onset of vibrational dynamics can therefore occur during the pulse itself. At the level of the KER spectrum, this effect is amplified by the fact that a small displacement of the nuclei can cause a rather large shift of the Coulomb explosion peak, due to the law of repulsion in $E=1/R$. For example, a Coulomb explosion occurring at $R=1.4$\;a.u. would produce a peak at 9.7\;eV, whereas at $R=1.7$\;a.u. the associated energy would be only 8.0\;eV. We can also see in panel \textbf{b} of Fig.\;\ref{Fig:One_Cycle} that the inclusion of vibrational dynamics results in a less intense Coulomb explosion peak at a slightly higher energy (8.9\;eV) than without vibrational dynamics (8.2\;eV). As we will see, this effect can be explained by the vibrational dynamics induced at short times.

In panel \textbf{d} of Fig.\;\ref{Fig:One_Cycle} we can observe that during the $800$\;nm pulse the population loss in \ch{H2} (black line) and the population gain in \ch{H2+} (red line) are strictly identical with and without vibrational dynamics. In contrast, the Coulomb explosion probability (green lines) decreases significantly with the inclusion of vibrational dynamics. This variation in the Coulomb explosion probabilities for identical \ch{H2+} populations is due to the fact that once \ch{H2+} is produced, a slight motion of the nuclei occurs. This can be seen in the inset within panel \textbf{d}. It shows the time evolution of the average internuclear distance $\langle R\,\rangle$ associated with the wave packet $\Psi_g(R,t)$ of \ch{H2+} between times $t=0.45\,t_f$ and $t=0.55\,t_f$, i.e. when Coulomb explosion occurs near the peak of the pulse. Single ionization of \ch{H2} initially produces a wave packet in \ch{H2+} around $\langle R\,\rangle \simeq 1.5$\;a.u. This ionization at an internuclear distance larger than the equilibrium distance of \ch{H2} (1.4\;a.u.) is due to the fact that the ionization potential decreases with the internuclear distance, and therefore the laser field ionizes more efficiently, and therefore more rapidly, the part of the \ch{H2} vibrational wave function located at $R > R_{eq} = 1.4$\;a.u. than the part located at a shorter internuclear distance, which ionizes more slowly. Thus, \ch{H2+} formation proceeds progressively, starting with higher internuclear distances and then with shorter distances. This time and R-dependent phenomenon induces some vibrational dynamics towards small inter-nuclear distances, which in turn leads to a temporal variation of the average internuclear distance in \ch{H2+}. In a separate calculation, we could verify that removing this $R$-dependence of the ionization rate suppresses the small initial compression of the chemical bond in \ch{H2+}. This motion toward shorter distances when vibrational dynamics is taken into account explains the shift of the Coulomb explosion peak toward higher energies. It also explains the lower probability of Coulomb explosion as the second ionization potential increases at shorter distances.

\begin{figure}[!t]
\centering
\includegraphics[height=10cm]{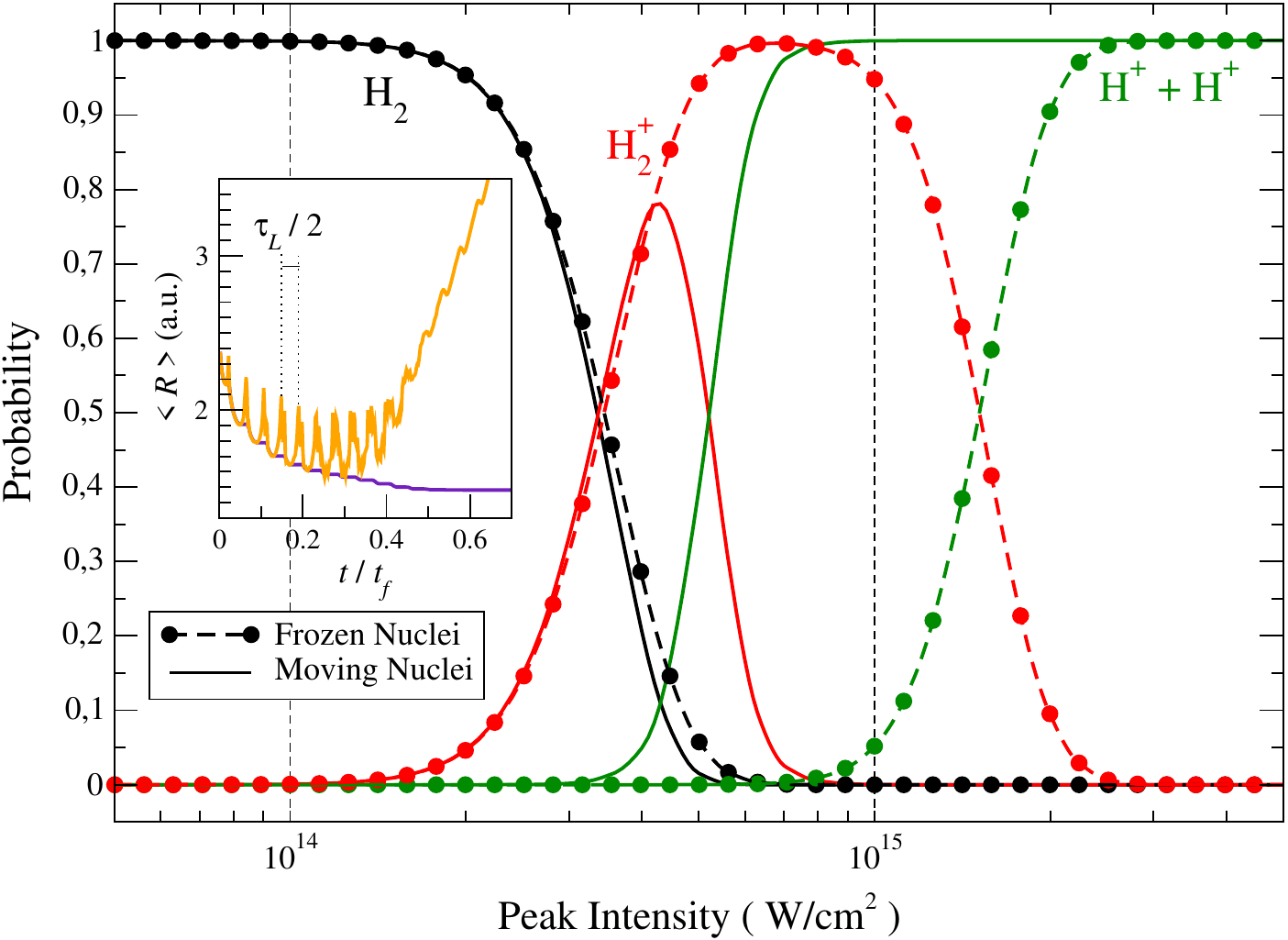}
\caption{Final populations in \ch{H2} (black lines), in \ch{H2+} (red lines), and in the Coulomb explosion channel \ch{H+} + \ch{H+} (green lines) as a function of the peak intensity (log scale between $5 \times 10^{13}$\;W/cm$^2$ and $5 \times 10^{15}$\;W/cm$^2$) using a 800\,nm linearly polarized field with a $\sin^2$ pulse envelope of total duration of 32\,fs, corresponding to 12 optical cycles. The dashed lines with circles show the results obtained by freezing the vibrational dynamics, while the solid lines take it into account. The inset shows the time evolution of the average internuclear distance $\langle R \rangle$ (in atomic units) of \ch{H2+} between $t=0$ and $t=0.7\,t_f$ for the peak intensity $6.3 \times 10^{14}$\, W/cm$^2$ in the case where the nuclear dynamics is frozen (lower decreasing purple line) and in the case where it is not (upper oscillating orange line).}
\label{Fig:Ionization_I}
\end{figure}

We now turn our attention to the double ionization process with pulses of several optical cycles. Fig.\;\ref{Fig:Ionization_I} shows the final populations (at $t=t_f$) in \ch{H2} (black lines), in \ch{H2+} (red lines) and in the Coulomb explosion channel \ch{H+} + \ch{H+} (green lines) as a function of the peak laser intensity in the case of a pulse of 12 optical cycles at 800\;nm (corresponding to a sin$^2$ envelope of total duration 32\,fs and full width at half maximum FWHM 16\,fs). The dashed lines with circles lines show the results obtained by freezing the vibrational dynamics, while the solid lines take it into account. The laser intensities considered range from $5 \times 10^{13}$\;W/cm$^2$ to $5 \times 10^{15}$\;W/cm$^2$, corresponding to maximum electric field amplitudes between 0.038 and 0.38\;a.u., thus approaching the electric field felt by an electron in a 1s orbital of the hydrogen atom, while remaining below this threshold where the laser field would completely dominate the attractive Coulomb field of the nuclei.

We can immediately see that the inclusion of vibrational dynamics has a very small, almost negligible effect on the final non-ionized population of \ch{H2}, whatever the laser intensity considered. In practice, \ch{H2} remains almost entirely in the ground vibrational state $v=0$ for the entire duration of the pulse, and the vibrational wave packet describing the \ch{H2} molecule (not shown here) evolves only by a gradual decrease of its norm. From all the simulations we have performed we can conclude that the neutral molecule \ch{H2} is not affected by any nuclear motion during the pulse, whatever the pulse intensity. By comparing the solid and dashed red lines in Fig.\;\ref{Fig:Ionization_I}, we can also note that when the laser intensity is too low to induce the removal of the second electron, i.e.\! for intensities below $4 \times 10^{14}$\;W/cm$^2$, the final \ch{H2+} molecular ion population is virtually unaffected by the inclusion of vibrational dynamics. At such intensities, ionization calculations performed at fixed inter-nuclear distances are therefore very realistic. In contrast, for laser intensities high enough to induce double ionization of the molecule, the final populations in \ch{H2+} and in the Coulomb explosion channel differ significantly with and without vibrational dynamics. This is because, once formed, the wave packet describing the \ch{H2+} molecular ion undergoes significant vibrational dynamics, leading it to explore internuclear distances greater than the distance at which it was formed. This is confirmed by the inset in Fig.\;\ref{Fig:Ionization_I}, which shows the time evolution of the average internuclear distance associated with \ch{H2+}, with and without vibrational dynamics. When the vibrational dynamics is frozen, the average internuclear distance at which \ch{H2+} is formed decreases progressively between the beginning of the pulse at $t=0$ and the pulse peak at $t=t_f/2$. This is because higher laser intensities allow the ionization of the molecule to occur at shorter internuclear distances, where the ionization potential is higher. We thus observe that initially \ch{H2+} is formed around $R = 2.4$\;a.u., but that from $t=t_f/2$ the average internuclear distance in \ch{H2+} stabilizes around $R = 1.5$\;a.u. The range of internuclear distances seen by \ch{H2+} is therefore, in this case, rather limited to short internuclear distances where double ionization is at a disadvantage due to a high second ionization potential. However, when the vibrational dynamics is taken into account, a completely different result is observed. In this case, at each half-optical-cycle, a burst of ionization occurs at short internuclear distances, followed by an evolution of \ch{H2+} towards larger internuclear distances. Then, half an optical cycle later, a new wave packet is again formed in \ch{H2+} at short internuclear distance, thus decreasing the average inter-nuclear distance in \ch{H2+}. This phenomenon then repeats itself at every half optical cycle, producing the oscillations shown in the inset. This is reflected in the evolution of the average internuclear distance $\langle R\, \rangle$ of \ch{H2+} by the appearance of a series of peaks separated by half an optical period ($\tau_L/2$, see the inset in Fig.\;\ref{Fig:Ionization_I}). Moreover, once the peak of the pulse is reached, single ionization begins to saturate and the mean internuclear distance in \ch{H2+} starts to increase almost linearly with time up to relatively high values, e.g. $R=3.1$\;a.u. at $t=0.6\,t_f$. By traveling toward large internuclear distances, the molecular ion \ch{H2+} explores regions where the ionization potential is lower than at short distances, allowing double ionization at relatively moderate laser intensities. This double ionization mechanism is obviously absent in a calculation that freezes the motion of the nuclei. This explains why the intensity threshold at which double ionization begins is higher when the vibrational dynamics are frozen.

\begin{figure}[!t]
\centering
\includegraphics[height=10cm]{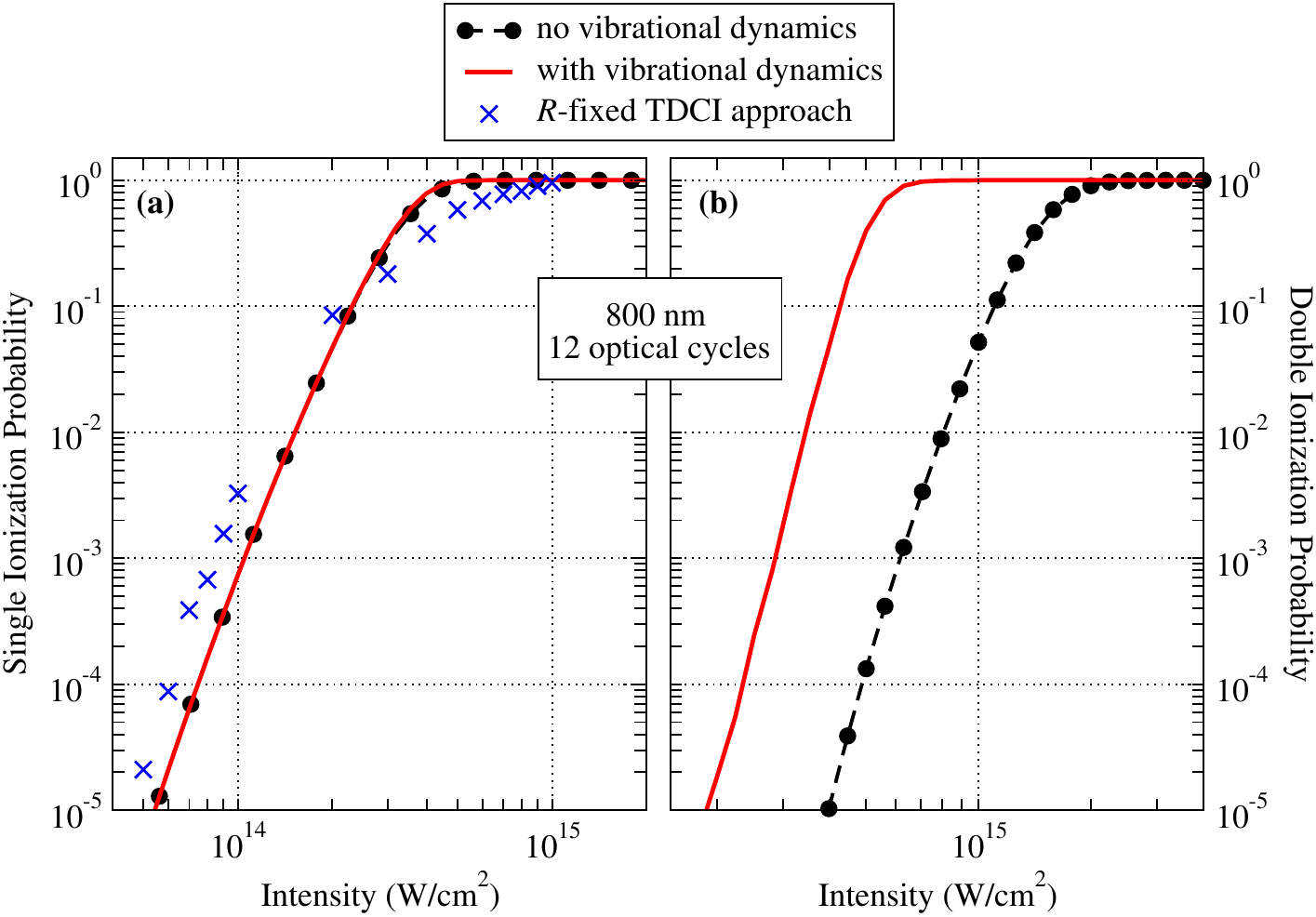}
\caption{Single (panel \textbf{a}) and double (panel \textbf{b}) ionization probabilities in a log-log scale as a function of the peak intensity between $4 \times 10^{13}$\;W/cm$^2$ and $2 \times 10^{15}$\;W/cm$^2$ (panel \textbf{a}) and between $1.5 \times 10^{14}$\;W/cm$^2$ and $4 \times 10^{15}$\;W/cm$^2$ (panel \textbf{b}) using a 800\,nm linearly polarized field with a $\sin^2$ pulse envelope of total duration of 32\,fs, corresponding to 12 optical cycles. The black dashed lines with circles show the results obtained by freezing the vibrational dynamics, while the solid red lines take it into account. The blue crosses were extracted from \citet{Awasthi2008} who used an $R-$fixed TDCI approach with the same laser parameters.}
\label{Fig:Ionization_12cycles}
\end{figure}

To highlight these differences on a larger intensity scale, Fig.\;\ref{Fig:Ionization_12cycles} shows the single and double ionization probabilities as a function of intensity on a logarithmic scale for the same laser parameters as in Fig.\;\ref{Fig:Ionization_I}. The black dashed lines with circles show the results obtained by freezing the vibrational dynamics, while the red solid lines take it into account. Panel \textbf{a} is devoted to the probability of single ionization, and panel \textbf{b} shows the probability of double ionization. The blue crosses show the single ionization probability obtained by a full Time-Dependent Configuration Interaction (TDCI) {calculation\cite{Vanne2004, Awasthi2005, Awasthi2008}}, in which the two-electron time-dependent Schr\"odinger equation was solved assuming that the vibrational motion is frozen and using CI wavefunctions consisting of thousands of configurations using Kohn-Sham (DFT) molecular orbitals expressed in an extended B-spline basis. The TDCI data used here were taken from\;\citet{Awasthi2008}. We see that our model including vibrational motion is in fairly good agreement with the full TDCI results obtained with fixed nuclei, confirming our earlier conclusion that the single ionization process can be safely described without considering vibrational dynamics. Finally, for the second ionization (panel \textbf{b} of Fig.\;\ref{Fig:Ionization_12cycles}), we can see that regardless of the laser intensity, taking into account the vibrational dynamics increases the probability of double ionization significantly, since in most cases the probability of double ionization is underestimated by several orders of magnitude when the nuclei are fixed.

\begin{figure}[!t]
\centering
\includegraphics[height=11cm]{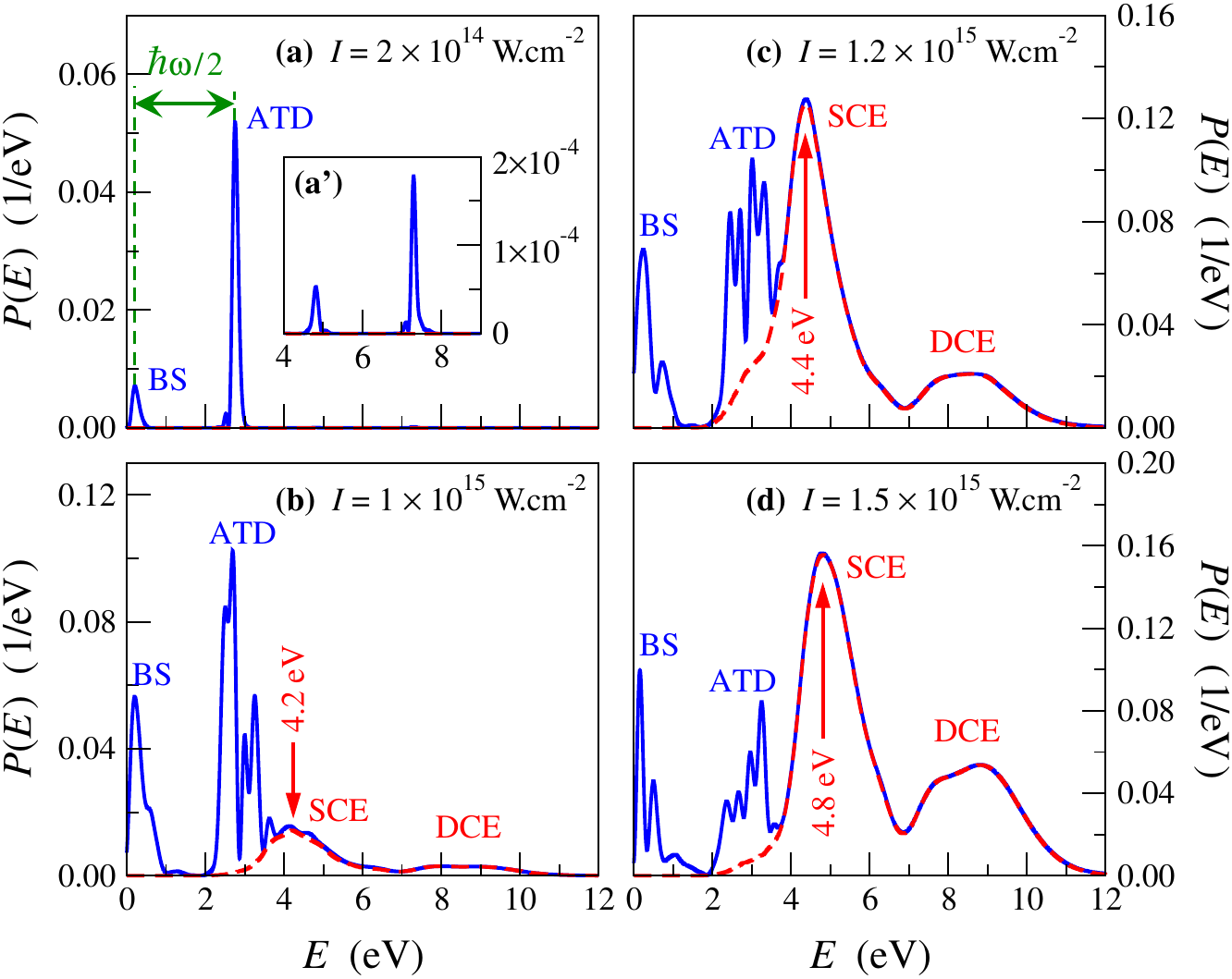}
\caption{Proton Kinetic Energy Release (KER) spectra calculated at 266\;nm for a peak intensity of \mbox{$2 \times 10^{14}$\;W/cm$^2$} (panels \textbf a and \textbf{a'}), \mbox{$10^{15}$\;W/cm$^2$} (panel \textbf b), \mbox{$1.2 \times 10^{15}$\;W/cm$^2$} (panel \textbf c) and \mbox{$1.5 \times 10^{15}$\;W/cm$^2$} \mbox{(panel \textbf d).} Panel \textbf{a'} is a simple zoom of panel \textbf a between 4 and 9\;eV. The electric field envelope is characterized by a sin$^2$ shape, and the total pulse duration is 32\;fs, corresponding to 36 optical cycles. The total KER spectra $P(E)$ are shown with blue solid lines, and the red dashed lines correspond to the KER spectra $P_{\mathrm{c}}(E)$ associated only with the Coulomb explosion channel. The total probabilities of photodissociation and Coulomb explosion are 1.2\,\% and 0\,\% in panel (\textbf a), 8.4\,\% and 3.6\,\% in panel (\textbf b), 10.4\,\% and 27.6\,\% in panel (\textbf c), and 8.8\,\% and 43.8\,\% in panel (\textbf d), respectively. The acronyms BS, ATD, SCE and DCE stand for Bond Softening, Above Threshold Dissociation, Sequential Coulomb Explosion and Direct Coulomb Explosion respectively.}
\label{Fig:Spectra}
\end{figure}

In this final section, we present characteristic proton Kinetic Energy Release (KER) spectra $P(E)$ predicted by our theoretical model at 266\;nm for different laser intensities. These spectra, given in Fig.\;\ref{Fig:Spectra}, are calculated for a total pulse duration of 32\;fs corresponding to 36 optical cycles. The total KER spectrum $P(E)$, i.e. including both the \ch{H+} + \ch{H}(1s) dissociation channel of the \ch{H2+} molecular ion and the \ch{H+} + \ch{H+} Coulomb explosion channel, is shown as a solid blue line, while the spectrum $P_{\mathrm{c}}(E)$ associated only with the Coulomb explosion channel is shown as a dashed red line.

\begin{figure}[!t]
\centering
\includegraphics[height=11cm]{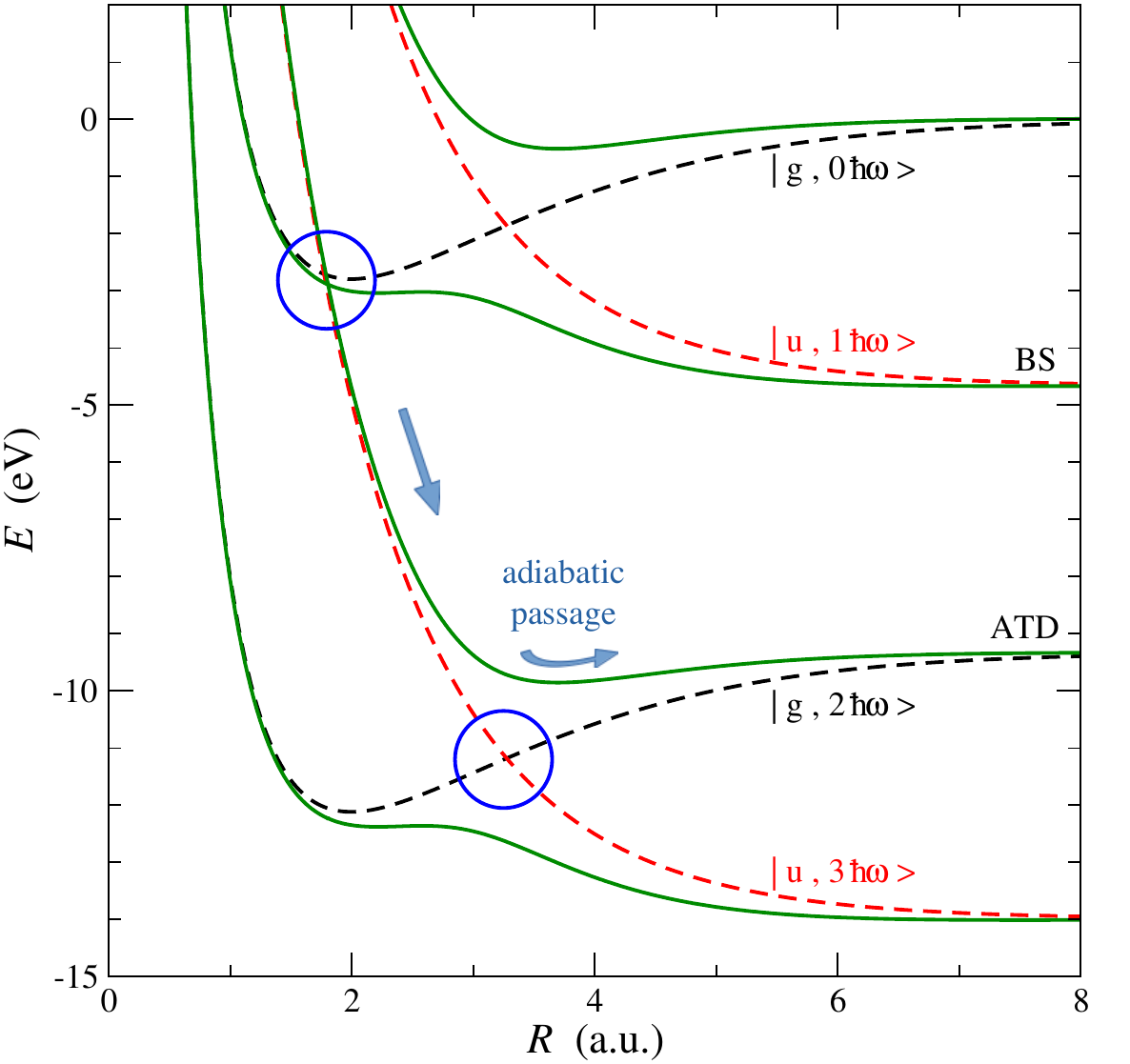}
\caption{\ch{H2+} 1s$\sigma_g$ (dashed black line) and 2p$\sigma_u$ (dashed red line) diabatic dressed potential curves for a radiation at 266\;nm. The adiabatic dressed potential curves are shown as solid green lines for the laser intensity $I = 2 \times 10^{14}$\;W/cm$^2$. The two most important potential curve crossings, which determine the photodissociation dynamics, are highlighted by the two blue circles. The asymptotic channels corresponding to Bond Softening and Above Threshold Dissociation are labeled by the acronyms BS and ATD.}
\label{Fig:Dressed_Curves}
\end{figure}

The KER spectra calculated for the peak intensity $2 \times 10^{14}$\;W/cm$^2$ is shown in the upper left panel \textbf{(a)}. At this relatively modest intensity we can see that there is no Coulomb explosion and that the entire spectrum corresponds to the photodissociation of the \ch{H2+} molecular ion. We can also see that this spectrum essentially consists of a series of two consecutive peaks, with the higher energy peak largely dominating the lower energy peak. This is characteristic of an above threshold dissociation (ATD) {process\cite{Giusti-Suzor1990, Bucksbaum1990, Zavriyev1990, Yang1991, Jolicard1992}}. This effect is  illustrated in Fig.\;\ref{Fig:Dressed_Curves}, which shows the field-dressed potential energy curves\cite{Bandrauk1981} of the \ch{H2+} molecule. At 266\,nm the potential curve of the electronic ground state 1s$\sigma_g$ (black dashed line $|\mathrm{g},0\hbar\omega\rangle$) crosses the potential curve of the 3-photon dressed excited state 2p$\sigma_u$ (red dashed line $|\mathrm{u},3\hbar\omega\rangle$) at an internuclear distance close to the equilibrium distance of 2\;a.u.\! of \ch{H2+}. The molecular ion \ch{H2+}, once formed by the ionization of \ch{H2}, readily absorbs 3 photons and begins to dissociate in the $|\mathrm{u},3\hbar\omega\rangle$ channel. The separating atomic fragments then pass through the avoided crossing between the $|\mathrm{g},2\hbar\omega\rangle$ and $|\mathrm{u},3\hbar\omega\rangle$ channels, around $R \simeq 3.3$\,a.u., where the molecule re-emits a quanta of photon energy to the radiation field by stimulated emission. The photodissociation of the molecular ion \ch{H2+} then proceeds adiabatically along the $|\mathrm{g},2\hbar\omega\rangle$ pathway, giving rise to the main peak of the KER spectrum at energy $E \simeq 2.7$\,eV. The peak observed at lower energies ($E \simeq 0.1$\,eV) is due to a bond softening\cite{Giusti-Suzor1990, Bucksbaum1990, Zavriyev1990, Yang1991, Jolicard1992} (BS) mechanism, which occurs when the potential barrier holding the bond is lowered sufficiently by the radiative interaction in the dressed potential curve picture that the molecule becomes unbound. The energy difference between these two peaks is half the photon energy (2.65\,eV), since each fragment (H or \ch{H+}) carries half the total energy released in the dissociation. The \textbf{(a')} panel within the \textbf{(a)} panel of Fig.\;\ref{Fig:Spectra} shows a zoom of the same spectrum over a higher energy range, between 4 and 9\;eV. Two additional peaks can be seen in this energy range, much less intense than the two main peaks already described. They correspond to the asymptotic dissociation channels $|\mathrm{u},3\hbar\omega\rangle$ and $|\mathrm{g},4\hbar\omega\rangle$. These two higher order peaks consist of a replica of the two peaks already described, which is consistent with the periodicity of the dressed potential curves.

As the laser intensity increases, double ionization occurs, changing the kinetic energy spectrum of the protons. This can be seen in the panel \textbf{(b)} of Fig.\;\ref{Fig:Spectra}, for the peak intensity $10^{15}$\,W/cm$^2$. In addition to the BS and ATD peaks, higher energy protons, mainly between 3 and 6\;eV, emerge from the ionization of \ch{H2+} at internuclear distances between 2.2 and 4.5\;u.a. This is therefore a sequential mechanism of Coulomb explosion of \ch{H2} via the intermediate step of formation and stretching of \ch{H2+}. The peak associated with this process is therefore labeled SCE for Sequential Coulomb Explosion in Fig.\;\ref{Fig:Spectra}. As observed {experimentally\cite{Saugout2007, Saugout2008}}, this peak grows and shifts to higher energies with increasing laser intensity, because the second ionization takes place at shorter internuclear distances for higher intensities. This effect is also observed in our calculations, with the SCE peak shifting from about 4.2\;eV at $10^{15}$\,W/cm$^2$ (panel \textbf{b}) to 4.4\;eV at $1.2 \times 10^{15}$\,W/cm$^2$ (panel \textbf{c}) and to 4.8\;eV at $1.5 \times 10^{15}$\,W/cm$^2$ \mbox{(panel \textbf{d})}. Finally, as the laser intensity increases, a second, higher energy Coulomb explosion peak is gradually formed, in the range of 7 to 10\;eV, corresponding to double ionization at very short internuclear distances of the order of 1 to 2\;a.u. This mechanism therefore corresponds to direct double ionization without the intermediate step of stretching of the molecular ion \ch{H2+}. This peak is therefore labeled DCE for Direct Coulomb Explosion in Fig.\;\ref{Fig:Spectra}.

\section{Conclusions}

In a previous work\cite{Vigneau2023} we have shown how well the PPT theory describes the single and double ionization rates of the \ch{H2} molecule for fixed nuclear geometries, considered as a parameter of the molecule, in a wide range of field frequency and intensity conditions covering both the tunnel and multiphoton ionization regimes. Capitalizing on this, we have developed an algorithm that combines the PPT description of the ionization steps with a quantum mechanical wave packet propagation procedure that takes into account the field-induced nuclear dynamics on all relevant channels during the ionization processes. Our calculations have highlighted the very important role played by nuclear dynamics in the \ch{H2+} molecular ion, with a strong influence on the double ionization of the molecule. Indeed, if the vibrational dynamics are not taken into account, the double ionization probabilities are generally underestimated by several orders of magnitude. We have also shown that this model gives realistic estimates of the ionization probabilities and allows one to easily calculate the kinetic energy of the protons emitted by photodissociation of \ch{H2+} through Above Threshold Dissociation and Bond Softening pathways and by Coulomb Explosion, thus unraveling the intensity dependent superposition of signals arising from these channels. This model can be used to obtain an approximate estimate of quantities of interest in the physics of laser-plasma interactions, where it could find some applications over a wide range of laser intensities, pulse durations, and wavelengths.

\bibliography{manuscript}

\end{document}